\documentclass[aps,twocolumn,groupedaddress,showpacs]{revtex4}
\usepackage{graphicx}
\begin{document}
%%%%%%%%%%%%%%%%%%%%%%%%%%%%%%%%%%%%%%%%%%%%%%%%%%%%%%%%%%%%%%%%%%%%%%%%%%%%%
% You should use BibTeX and revtex.bst for references
%\bibliographystyle{apsrev}
%%%%%%%%%%%%%%%%%%%%%%%%%%%%%%%%%%%%%%%%%%%%%%%%%%%%%%%%%%%%%%%%%%%%%%%%%%%%%
% marks overfull lines with blackboxes
%\draft - no longer supported, use the 'draft' option instead
%\preprint{}
\title{Magnetic properties of the spin-$\frac{1}{2}$ XXZ model on the Shastry-Sutherland lattice: Effect of long-range interactions}
% Optional argument for running titles on pages
%\title[]{}
\author{Takahumi Suzuki, Yusuke Tomita and Naoki Kawashima}
%\thanks{}
\affiliation{Institute for Solid State Physics, University of Tokyo, Kashiwa, Chiba 277-8581, Japan}
\date{\today}
%%%%%%%%%%%%%%%%%%%%%%%%%%%%%%%%%%%%%%%%%%%%%%%%%%%%%%%%%%%%%%%%%%%%%%%%%%%%%
%                         ABSTRACT                                          %
%%%%%%%%%%%%%%%%%%%%%%%%%%%%%%%%%%%%%%%%%%%%%%%%%%%%%%%%%%%%%%%%%%%%%%%%%%%%%
\begin{abstract}
We study magnetic properties of the $S=1/2$ Ising-like XXZ model on the Shastry-Sutherland lattices with long-range interactions, using the quantum Monte Carlo method. This model shows magnetization plateau phases at one-half and one-third of the saturation magnetization when additional couplings are considered. We investigate the finite temperature transition to one-half and one-third plateau phases. The obtained results suggest that the former case is of the first order and the latter case is of the second order. We also find that the system undergoes two successive transitions with the 2D Ising model universality, although there is a single phase transition in the Ising limit case. Finally, we estimate the coupling ratio to explain the magnetization process observed in ${\rm TmB_4}$. 

\end{abstract}
%%%%%%%%%%%%%%%%%%%%%%%%%%%%%%%%%%%%%%%%%%%%%%%%%%%%%%%%%%%%%%%%%%%%%%%%%%%%%
%\maketitle must follow title, authors, abstract and \pacs
\maketitle
% References should be done using the \cite, \cite, and \label commands
%\section{}
%\label{}
%\subsection{}
%\subsubsection{}
%%%%%%%%%%%%%%%%%%%%%%%%%%%%%%%%%%%%%%%%%%%%%%%%%%%%%%%%%%%%%%%%%%%%%%%%%%%%%
%                        MAIN TEXT                                          %
%%%%%%%%%%%%%%%%%%%%%%%%%%%%%%%%%%%%%%%%%%%%%%%%%%%%%%%%%%%%%%%%%%%%%%%%%%%%%
%\section{INTRODUCTION}
%%%%%%%%%%%%%%%%%%%%%%%%%%%%%%%%%%%%%%%%%%%%%%%%%%%%%%%%%%%%%%%%%%%%%%%%%%%
Magnetic properties of quantum spin systems with frustrated antiferromagnetic interactions have been an interesting topic from both theoretical and experimental aspects, because the strong fluctuations prevent the stabilization of classical orderings. 
Some exotic ordered states, such as a spin supersolid state on the triangular lattice\cite{supersolid2,supersolid4}, a $Z_2$ spin liquid on the kagome lattice\cite{kagome1,kagome2}, and a $U(1)$ liquid state on the pyrochlore lattice\cite{U1}, are good examples reflecting such fluctuations. 
The $S=1/2$ antiferromagnetic Heisenberg model on the Shastry-Sutherland lattices (SSLs)\cite{SSL1,SrCuBO0,SrCuBO1,Ueda} has been studied as the other class with interesting characteristics derived from frustration and quantum fluctuation. 
In previous studies, it was suggested that this model has rich magnetization phases including spin supersolid phases at moderate fillings between magnetization plateau phases\cite{momoi}. 
From experimental observations for the SSL compound $\rm{SrCu(BO_3)_2}$\cite{SrCuBO0}, it was also discussed that there is the possibility of fractionalized magnetic plateaus at the magnetization $m$=$m_z/m_s$=$1/5, 1/6, 1/7, 1/9$ and $2/9$ in addition to the previously reported plateaus at $m$=$1/3$ and $1/4$\cite{SrCuBO1}.
Here, $m_s$ indicates the saturation magnetization.
While one theoretical scenario for explaining these plateaus was proposed in analogy to the quantum Hall effect\cite{SrCuBO1}, a few points, such as the number of magnetization plateaus and the magnetization values of plateaus, are at the moment controversial.

In recent experiments on rare-earth metals $\rm{RB_4}$ [R is a rare-earth atom]\cite{TbB4,ErB4,TmB4_1,TmB4_2,Siemensmeyer}, fractionalized magnetic plateaus were also discovered at a very low temperature.
These compounds have a tetragonal crystal structure $\rm{P4/mbm}$ and the magnetic moments originating from ${\rm R^{3+}}$ locate on the SSL in the $ab$-plane. 
In ${\rm TmB_4}$, a large $m$=$1/2$-magnetization-plateau region was confirmed for 1.9[T]$<H<$3.6[T] at a low temperature when the magnetic fields are applied along the $c$-axis\cite{TmB4_1,TmB4_2}. 
An important point that is the most different from $\rm{SrCu(BO_3)_2}$ is that, in the ${\rm TmB_4}$ case, a strong anisotropy along the $c$-axis is expected owing to the crystal fields.
Recent analysies for specific heat measurements and the magnetization process\cite{TmB4_2} suggested that the $J$=$6$ multiplet of ${\rm Tm^{3+}}$ is lifted, and then, the lowest energy state for a single ion is the non-Kramers doublet with $J_z$=$\pm 6$, having a large energy gap between the lowest states.
This leads us to describe the low-energy magnetic part of ${\rm TmB_4}$ by a binary model. Since it is expected that isotropic interactions work between the moments via itinerant electrons, we start from the two-site Hamiltonian, $H_0+H'$, where $H_0$=$-D\sum_i (J_i^z)^2$, $H'$=$G\sum_{ij}{\bf J}_i \cdot {\bf J}_j$, and $0$$<$$|G|$$\ll$$D$. In the ground state of $H_0$, there are four degenerated states, namely $|J_i^z,J_j^z \rangle$=$|\pm 6,\pm 6\rangle$ and $|\pm 6,\mp 6\rangle$. When we treat $H'$ as the perturbation, the first nontrivial off-diagonal term occurs from the 2$J$-th order. The obtained effective Hamiltonian is the $S=1/2$ Ising-like XXZ Hamiltonian with the transverse term proportionally to $-\{G/(-D)\}^{2J}$. Note that the transverse term always becomes $ferromagnetic$ independently on the sign of $G$ because $J$ is integer.
Meng and Wessel studied the magnetization curves for the effective Hamiltonian by quantum Monte Carlo computations\cite{TmB4_4}. From the obtained results, it was found that the origin of the $m$=$1/2$ plateau can be explained by the quantum effect because only the $m$=$1/3$ plateau was confirmed in the Ising model case. 
However, the phase diagram for the Ising-like XXZ model case showed that the $m$=$1/3$ plateau phase spreads more widely than the $m$=$1/2$ plateau phase. 
Since the $m$=$1/3$ plateau was not observed in $\rm{TmB_4}$, their results seem to be inconsistent with the experimental observations. 
This inconsistency may arise from the absence of ``long-rang'' interactions among magnetic moments of $\rm{Tm^{3+}}$ ions, namely, the RKKY interactions.  
The magnetic properties of the Ising-like XXZ model with the long-range interactions have not studied yet.
Thus, the clarification of the effects of long-range interactions on the magnetic properties is an important problem and helps us understand the origin of the $m$=$1/2$ plateau observed in ${\rm TmB_4}$.

\begin{figure}[bth]
  \begin{center}
  \includegraphics[scale=0.5]{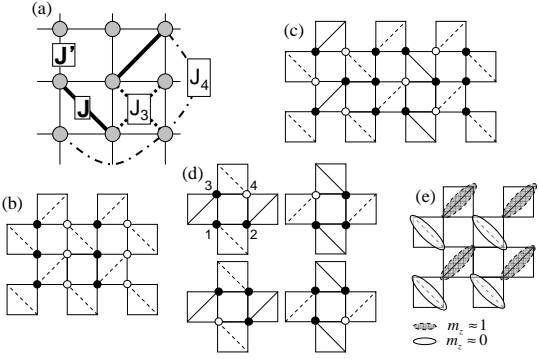}
  \end{center}
  \vspace*{-7mm}
  \caption{(a) Effective model on the SSL with diagonal coupling $J$, nearest-neighbor coupling $J'$, and additional couplings $J_3$ and $J_4$. The $J_3$ coupling is defined on every plaquette without diagonal coupling, and the $J_4$ coupling corresponds to the third-nearest neighbor coupling of the orthogonal-dimer-lattice geometry. (b), (c), and (d) are typical spin configurations in the $m=0$ collinear, $1/3$ plateau, and $1/2$ plateau states, respectively. In (b)$\sim$(d), solid (open) circles represent up spins (down spins) and the broken (solid) line on each diagonal bond indicates an antiparallel (parallel) spin pair. (e) shows the bond N\'eel state in the intermediate phase (see text).}
\label{spin_config}
\end{figure}

In this paper, we discuss the magnetic properties of the $S=1/2$ Ising-like XXZ model on the SSL using the quantum Monte Carlo method based on the modified directed-loop algorithm\cite{KatoYasu}. We consider the effects of $J_3$ and $J_4$ couplings in addition to the conventional SSL model with $J$ and $J'$ couplings. The geometrical configuration of the additional couplings $J_3$ and $J_4$ is shown in Fig. \ref{spin_config} (a).
The Hamiltonian considered here is described by 
%%%%%%%%%%%%%%%%%%%%%%%%%%%%%%%%%%%%%%%%%%%%%%%%%%%%%%%%%%%%%%%%%%%%%%%%%%%
\begin{eqnarray}
{\mathcal H}&=&\sum_{\langle i,j \rangle} \left( J {\bf S}_{i}\cdot{\bf S}_{j} \right)_{\Delta_\perp}+\sum_{\langle i,j \rangle'}\left( J'{\bf S}_{i}\cdot{\bf S}_{j} \right)_{\Delta_\perp}\nonumber\\&+&\sum_{\langle i,j \rangle''}\left( J_3{\bf S}_{i}\cdot{\bf S}_{j} \right)_{\Delta_\perp}
+\sum_{\langle i,j \rangle'''}\left( J_4{\bf S}_{i}\cdot{\bf S}_{j} \right)_{\Delta_\perp}\nonumber\\
&-&g\mu_B H\sum_{i}{S_{i}}^{z},
\label{Ham}
\end{eqnarray}
where $\left( J{\bf S}_{i}\cdot{\bf S}_{j} \right)_{\Delta_\perp}=-{\Delta}_{\perp}|J|\left( S_i^{x}S_j^{x}+S_i^{y}S_j^{y} \right) +J S_i^{z}S_j^{z}$ and $\langle ij \rangle$, $\langle ij \rangle'$,  $\langle ij \rangle''$, and $\langle ij \rangle'''$ mean sums over all pairs on the bonds with the $J$, $J'$, $J_3$, and $J_4$ couplings, respectively. Note that the positive (negative) sign of each coupling means an antiferromagnetic (ferromagnetic) interaction and $J>0$ and $J'>0$ are considered.
In the following computations, we set $g\mu_B=k_B=1$ and treat the $L\times L$ system with the periodic boundary condition.

 \begin{figure}[bth]
%  \begin{center}
  \includegraphics[scale=0.48]{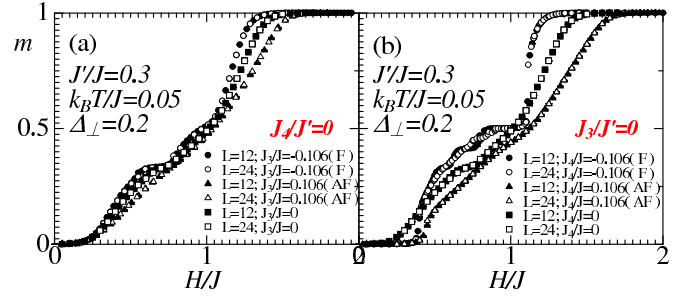}
%  \end{center}
  \vspace*{-7mm}                                                        
  \caption{Magnetization processes at $k_BT/J=0.05$. (a) and (b) are results when we consider $J_3$ and $J_4$ couplings, respectively. Circles (triangles) indicate the results for the additional ferromagnetic (antiferromagnetic) couplings. The magnetization processes without the additional couplings are represented by squares. The symbols are drawn with error bars, which are smaller than the symbol size (here and the following figures).}
\label{magnetization}
\end{figure}

\begin{figure}[bth]
  \begin{center}
  \includegraphics[scale=0.4]{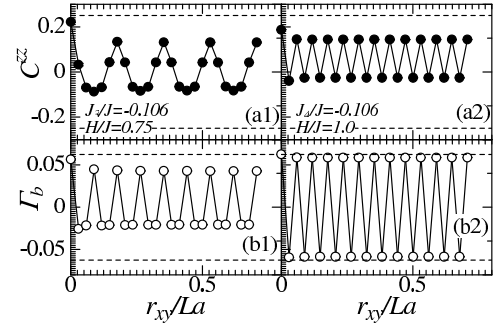}
  \end{center}
  \vspace*{-7mm}
  \caption{Bare spin and bond spin correlations in $1/2$ and $1/3$ plateau states at $k_BT/J=0.05$. The left-hand figures (a1) and (b1) show the results for the $1/3$ plateau state at $J_3/J=-0.106$ and $H/J=0.75$. The right-hand figures (a2) and (b2) show the results for the $1/2$ plateau state at $J_4/J=-0.106$ and $H/J=1$.  The horizontal axis is the distance along the (1,1) direction, which is normalized by the linear dimension $L=48$.}
\label{correlation}
\end{figure}

We present the obtained magnetization processes in Fig. \ref{magnetization} when the easy-axis anisotropy and the coupling ratio are fixed at ${\Delta}_{\perp}$=0.2 and $J'/J$=0.3, respectively. 
The magnetization-plateau states become stable at $m=1/2$ ($m=1/3$) for 1.85$<H/J<$2.2 (1.25$<H/J<$1.55) when the weak ferromagnetic $J_4$ ($J_3$) coupling is taken into account. 
To investigate the spin configurations in both plateau states, we calculate two types of correlation functions. One is the bare spin correlation defined by $C^{zz}({\bf r})=\langle S^z({\bf r})S^z(0)\rangle -\langle m_z\rangle^2$, and the other is the bond spin correlation defined by $\Gamma_b({\it\bf r}_d)=\langle B({\it\bf r}_d)B(0)\rangle-\langle B({\it\bf r}_d)\rangle^2$, where $B({\it{\bf r}}_d)={S^z}_{\it{\bf r}_d+\delta}{S^z}_{\it{\bf r}_d-\delta}$ and $B({\bf r}_d)$ is defined only on the plaquettes with the diagonal coupling $J$. 
The bond spin correlation is convenient for characterizing the crystallization of the triplet dimers with $S^z=1$ and $S^z=0$, which cannot be detected from the bare spin correlation.
From Fig. \ref{correlation}, we find that both spin correlations show the presence of a true long-range order at $m=1/3$ and $m=1/2$.
In the $L\times L$ lattice systems at $k_BT/J=0.05$, we confirm that no other plateau survives in the thermodynamic limit and that the ferromagnetic $J_4$ coupling tends to stabilize $only$ the 1/2 plateau even when the coupling ratio $J'/J$ varies. Thus, we believe that the ferromagnetic $J_4$ coupling plays a significant role in explaining the $1/2$ plateau observed in ${\rm TmB_4}$.

To discuss the reason why the field range of the $1/3$ or $1/2$ plateau is expanded by the ferromagnetic $J_3$ or $J_4$ coupling, it is constructive to consider the Ising limit case.
The lowest energy having the magnetization $m=0$, $1/3$, and $1/2$ can easily be estimated from each spin configurations at $T=0$. 
Fig. \ref{spin_config} shows the conventional spin configurations at a very low temperature, which are obtained from computations for the Ising model case.
At $m=0$, the collinear order shown in Fig. \ref{spin_config} (b) is stabilized by the $J_3$ and $J_4$ couplings, and the local energy is given by $\epsilon_{co}=-J/8-J_3/4+J_4/2$. In the same manner, we obtain $\epsilon_{1/3}=-J/24-J'/6+J_3/12+J_4/6-H/6$ for the $1/3$ plateau state and $\epsilon_{1/2}=J_4/2-H/4$ for the $1/2$ plateau state.
Since the energy of the fully polarized state is given by $\epsilon_{F}=J/8+J'/2+J_3/4+J_4/2-H/2$, the field ranges of the $1/3$ and $1/2$ plateaus can be calculated as $\Delta H_{1/3}=3J'-3J_3+6J_4$ and $\Delta H_{1/2}=2J_3-4J_4$ respectively.
Consequently the ferromagnetic $J_4$ ($J_3$) coupling makes the $1/2$ ($1/3$) plateau state stable, which is in contrast to the antiferromagnetic case.

\begin{figure}[bth]
  \begin{center}
  \includegraphics[scale=0.5]{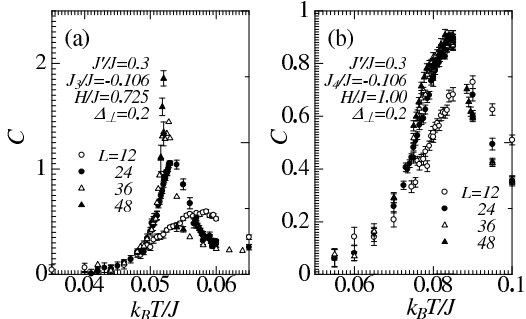}
  \end{center}
  \vspace*{-7mm}
  \caption{Temperature dependence of specific heat at (a) $(J_3/J,J_4/J,H/J)=(-0.106,0,0.725)$ and (b) $(J_3/J,J_4/J,H/J)=(0,-0.106,1.0)$.}
\label{Temp_spc}
\end{figure}

Next, we investigate the finite temperature transition to the plateau phases.
To assess the nature of the finite temperature transition to the $1/3$ and $1/2$ plateau states, we calculate the temperature dependence of the specific heat, and the results are shown in Fig. \ref{Temp_spc}. The specific heat at $(J_3/J,J_4/J,H/J)=(-0.106,0,0.725)$ has a single peak and the maximum value strongly diverges around $T/J\sim0.052$ for increasing the system size $L$. This implies that the phase transition from the paramagnetic phase to the $1/3$ plateau phase is of the first order. On the other hand, the weak system size dependence in the model at $(J_3/J,J_4/J,H/J)=(0,-0.106,1.0)$ seems to result from the second-order phase transition with a negative exponent $\alpha<0$.

\begin{figure}[bth]
  \begin{center}
  \includegraphics[scale=0.53]{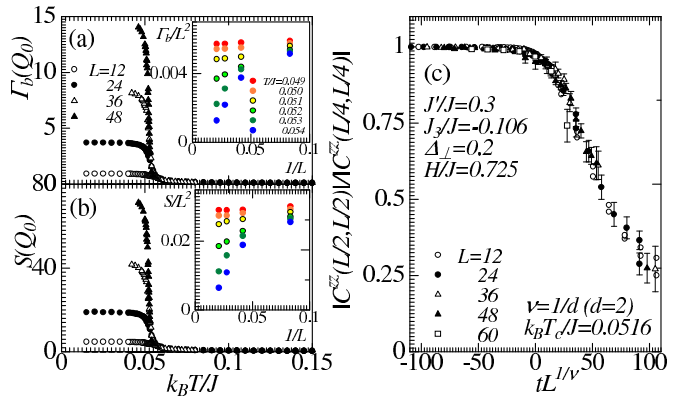}
  \end{center}
  \vspace*{-4mm}
  \caption{(Color online) Transition to $1/3$ plateau phase. Temperature dependence of static structure factors for (a) bare and (b) bond spins at $Q_0=(\pi/3,0)$. Insets show their size dependence. (c) Finite size scaling for correlation ratio of bond spins.}
\label{FSS_first}
\end{figure}

To clarify the order of the finite temperature transition to the $1/3$ plateau phase, we calculate static structure factors both for bare spins $S({\bf Q_0})$ and for bond spins $\Gamma_b({\bf Q_0})$ at ${\bf Q_0}=(\pi/3,0)$. The results are shown in Fig. \ref{FSS_first}. Both structure factors suddenly develop at $T/J\sim 0.052$ and the values increase proportionally to the system size below the onset temperature. This result also indicates the presence of the first-order transition.
In the phenomenological renormalization-group treatment for the first-order transition\cite{firstorder}, the bond-spin correlation ratio around the critical temperature scales as $\Gamma_b(L/2,L/2)/\Gamma_b(L/4,L/4)\sim F(tL^{1/\nu})$, where $F$ is a scaling function, $t=(T-T_c)/T_c$, and the critical exponent $\nu$ is given by the inverse of the space dimensionality $1/d (d=2)$. From Fig. \ref{FSS_first} (c), we find that the finite-size scaling analysis is successfully performed and the data collapse is confirmed at $T_c/J=0.516(3)$.

We continue to study the universality class of the phase transition to the $1/2$ plateau. From the low-temperature spin configuration in the Ising limit case (see Fig. \ref{spin_config} (d)), we can expect $C_4$ symmetry breaking. Since the lowest energy state of the $1/2$ plateau phase is fourfold degenerated in context of the bare spin language, the universality class of this phase transition is naively expected to be the four-state-Potts universality class (the critical exponents are given by $\nu=2/3$ and $\beta=1/12$). In fact, we confirm a single phase transition belonging to the four-state Potts universality class in the Ising limit case\cite{suzuki2}. However, it is also possible that the $C_4$ symmetry breaks down in two separate steps as we see below.

\begin{figure}[bth]
  \begin{center}
  \vspace*{2mm}
  \includegraphics[scale=0.5]{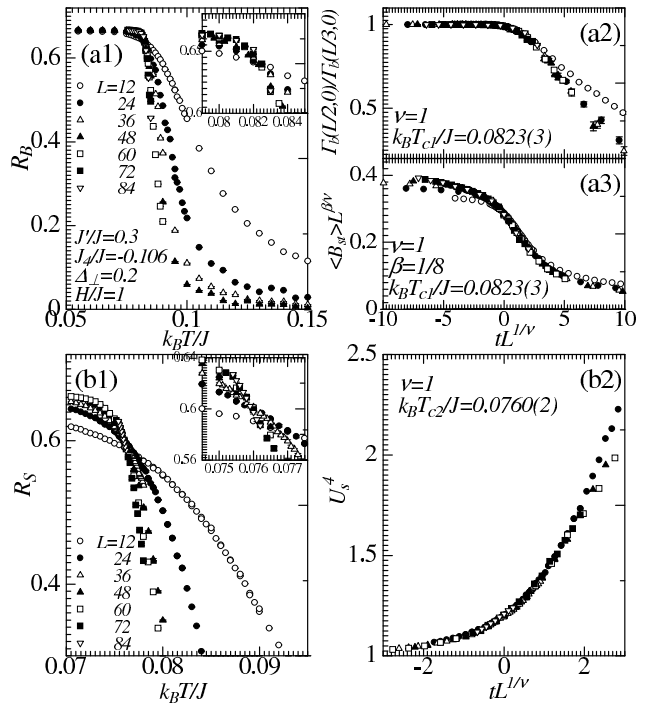}
  \end{center}
  \vspace*{-7mm}
  \caption{Transition to $1/2$ plateau phase. (a1) Temperature dependence of the Binder ratio $R_B$. Finite-size scaling analysis for (a2) correlation ratio and (a3) staggered magnetization $B_{st}$. (b1) Temperature dependence of the Binder ratio $R_S$ and (b2) finite size scaling for fourth-order cumulant $U_S^4$}
\label{FSS_second}
\end{figure}

By introducing the 90 degree lattice rotation ``$c_4$'' around the center of a plaquette without diagonal coupling, the symmetry group can be expressed as $C_4=\{e,c_4,c_4^2,c_4^{-1}\}$. At the higher critical temperature $T_{c1}$, the symmetry breaks down to $\{e,c_4^2\}$, which is the symmetry of the intermediate phase, and at the lower critical temperature $T_{c2}$, the remaining symmetry breaks down to the trivial group $\{e\}$.
To detect them, we introduce two kinds of order parameters here.  
The symmetry breaking at $T_{c1}$ can be characterized by a bond-spin ``staggered magnetization'' represented by $B_{st}=|\sum_{{\bf r}_d} (-1)^{f({\bf r}_d)}B({\bf r}_d)|$, where $f({\bf r}_d)$ takes $\pm 1$ depending on the position of the diagonal coupling. The other symmetry breaking at $T_{c2}$ can be characterized by freezing the antiferromagnetic fluctuation on the diagonal bond $J$, so that the order parameter is given by $m_x^c=S^z_4-S^z_1$ or $m_y^c=S^z_3-S^z_2$. Here, the suffixes of longitudinal spin operators represent the site indexes shown in Fig. \ref{spin_config} (d). Fig. \ref{FSS_second} (a1) shows the temperature dependence of the Binder ratio $R_B$ for the staggered magnetization $B_{st}$. We find that $R_B$ has a single crossing point at $k_BT_{c1}/J=0.0822(3)$. Since the $Z_2$ symmetry breaking at $T_{c1}$ is expected from the viewpoint of bond spin ordering, an intuitive speculation for the universality class of the phase transition leads us to expect the 2D Ising universality class. We thus perform the data collapse assuming the scaling form $\Gamma_b(L/2,0)/\Gamma_b(L/3,0)=F(tL^{1/\nu})$ and $B_{st}L^{\beta/\nu}=F(tL^{1/\nu})$ with $\nu=1$ and $\beta=1/8$. The results are shown in Figs. \ref{FSS_second} (a2) and (a3). The data collapse is confirmed with good accuracy despite the fact that we did not vary both the critical exponents and the critical temperature in the analysis. The temperature dependence of the Binder ratio $R_S$ of $m_x^c$ provides a clear evidence of another phase transition. As shown in Fig. \ref{FSS_second} (b1), $R_S$ has a single crossing point at $k_BT_{c2}/J=0.0760(2)$ and the value is clearly different from that of the bond-spin ordering temperature. The universality class of the phase transition at $T_{c2}$ is expected to be the same as that of the 2D Ising model because the remaining symmetry $C_2$ is isomorphic to $Z_2$. Actually, the finite-size scaling analysis for the fourth order cumulant $U_S^4$ of $m_x^c$ can be achieved by using $U_S^4=F(tL^{1/\nu})$ and $\nu=1$, as shown in Fig. \ref{FSS_second} (b2). From the obtained results, we conclude that there exists an intermediate phase that can be characterized by the bond N\'eel orderings accompanying the internal antiferromagnetic bare-spin fluctuation (see Fig. \ref{spin_config} (e)).

The obtained result has an interesting nature; the criticality is different from that of the four-state Potts model. One of the interpretations for this discrepancy is given by the phase diagram of the generalized four-state clock model, which may be similar to that of the generalized Ashkin-Teller model\cite{Nijs,AT}. In the Ising limit case, the phase transition corresponds to the $P_4$ fixed point on the self-dual line. If the transverse coupling $(S^+_iS^-_j+ h.c.)$ is added, the criticality may change owing to the breaking of self duality; the model undergoes two successive phase transitions with the 2D Ising universality or only one phase transition with the weak 2D Ising universality. Further investigations in this direction are now in progress and will be published elsewhere\cite{suzuki2}.

Finally, we discuss the magnetization of ${\rm TmB_4}$.
It was observed that the $1/2$ plateau state appears for 1.9[T]$<H<$3.6[T] at 2[K]\cite{TmB4_1}, and the observed magnetization process associates with the coupling ratio $J'/J \sim 1$ and the strong Ising anisotropy $\Delta_{\perp}<<1$\cite{Siemensmeyer}.
From these facts, we roughly estimate the coupling ratios $J_3/J$ and $J_4/J$ by the local energy estimation in the Ising model case with $J'/J=1$. We obtain the coupling ratios as $J_3/J\sim0.1182$ and $J_4/J<-0.25$ at $J'/J=1$ with $J\sim0.090$[K]\cite{comment}.
Note that the field range of the $1/2$ plateau state does not depend on $J_4$, although the $J_4$ coupling is important for the appearance of the $1/2$ plateau.
Furthermore we comment that this parameter set also gives a consistent N\'eel temperature $T_N \sim 7$[K] at $H=0$ (experimentally, it was $T_N\sim$ 9.6[K]).
In previous studies\cite{Siemensmeyer,TmB4_4}, it was clarified by the Monte Carlo calculations for the minimal Ising model with $J_3=J_4=0$ that the plateaus except the $1/3$ plateau are not stabilized\cite{TmB4_4}. 
Thus, our estimation seems to be reasonable.
However, checking its validity more precisely via comparison with the other experimental observations might be required, because a small plateau at $m \sim 1/8$, which our model couldn't replicate it, has been observed experimentally.

%In summary, we investigate the magnetic properties of the $S=1/2$ XXZ model on the SSL. The $1/3$ and $1/2$ plateaus are stabilized by the ferromagnetic $J_3$ and $J_4$ couplings, respectively. The finite temperature transition to the $1/3$ plateau phase is of the first order. When the finite temperature transition to the $m=1/2$ plateau phase takes places, the transverse coupling two successive phase transitions with the 2D Ising universality class are confirmed. 

%%%%%%%%%%%%%%%%%%%%%%%%%%%%%%%%%%%%%%%%%%%%%%%%%%%%%%%%%%%%%%%%%%%%%%%%%%%
\section*{Acknowledgments} 
%%%%%%%%%%%%%%%%%%%%%%%%%%%%%%%%%%%%%%%%%%%%%%%%%%%%%%%%%%%%%%%%%%%%%%%%%%%
The present research subject was suggested by C. D. Batista. We would like to thank him and P. Sengupta for fruitful communications. The computation in the present work is executed on computers at the Supercomputer Center, Institute for Solid State Physics, University of Tokyo. The present work is financially supported by Grant-in-Aid for Young Scientists (B) (21740245), Grant-in-Aid for Scientific Research (B) (19340109), Grant-in-Aid for Scientific Research on Priority Areas ``Novel States of Matter Induced by Frustration'' (19052004), and by Next Generation Supercomputing Project, Nanoscience Program, MEXT, Japan.

%%%%%%%%%%%%%%%%%%%%%%%%%%%%%%%%%%%%%%%%%%%%%%%%%%%%%%%%%%%%%%%%%%%%%%%%%%%%%
%                            REFERENCES                                     %
%%%%%%%%%%%%%%%%%%%%%%%%%%%%%%%%%%%%%%%%%%%%%%%%%%%%%%%%%%%%%%%%%%%%%%%%%%%%%
% Create the reference section using BibTeX
%\bibliography{nmr}

\begin{references}
%
\bibitem{supersolid2} S. Wessel and M. Troyer, Phys. Rev. Lett. {\bf 95}, 127205 (2005).
\bibitem{supersolid4} M. Boninsegni and N. Prokofev, Phys. Rev. Lett. {\bf 95}, 237204 (2005).
\bibitem{kagome1} L. Balents, M. P. A. Fisher, and S. M. Girvin, Phys. Rev. B {\bf 65}, 224412 (2002).
\bibitem{kagome2} D. N. Sheng and L. Balents, Phys. Rev. Lett. {\bf 94}, 146805 (2005).
\bibitem{U1} A. Banerjee, S. V. Isakov, K. Damle, and Y. B. Kim, Phys. Rev. Lett. {\bf 100}, 047208 (2008).
\bibitem{SSL1}  B. S. Shastry and B. Sutherland, Physica B {\bf 108}, 1069 (1981).
\bibitem{Ueda}  S. Miyahara and K. Ueda, J. Phys.: Condens. Matt. {\bf 15}, R327 (2003).
\bibitem{SrCuBO0} H. Kageyama, K. Yoshimura, R. Stern, N. V. Mushnikov, K. Onizuka, M. Kato, K. Kosuge, C. P. Slichter, T. Goto, and Y. Ueda, Phys. Rev. Lett. {\bf 82}, 3168 (1999).
\bibitem{SrCuBO1} S.E. Sebastian, N. Harrison, P. Sengupta, C. D. Batista, S. Francoual, E. Palm, T. Murphy, H. A. Dabkowska, and B. D. Gaulin, PNAS {\bf 105}, 20157 (2008).
\bibitem{momoi} T. Momoi and K. Totsuka, Phys. Rev. B {\bf 62}, 15067 (2000).
\bibitem{TbB4} S. Yoshii, T. Yamamoto, M. Hagiwara, T. Takeuchi, A. Shigekawa, S. Michimura, F. Iga, T. Takabatake, and K. Kindo, J. Magn. Magn. Mater. {\bf 310}, 1282-1284 (2007).
\bibitem{ErB4} S. Michimura, A. Shigekawa, F. Iga, M. Sera, T. Takabatake, K. Ohyama, and Y. Okabe, Physica B {\bf 378-380}, 596-597 (2006).
\bibitem{TmB4_1} S. Yoshii, T. Yamamoto, M. Hagiwara, A. Shigekawa, S.Michimura, F. Iga, T. Takabatake, and K. Kindo, J. Phys.: Conf. Ser. {\bf 51}, 59-62 (2006).
\bibitem{TmB4_2} S. Gab\'ani S. Matas, P. Priputen, K. Flachbart, K. Siemensmeyer, E. Wulf, A. Evdokimova, and N. Shitsevalova, Acta. Phys. Pol. A {\bf 116}, 227 (2008).
\bibitem{Siemensmeyer} K. Siemensmeyer, E. Wulf, H.-J. Mikeska, K. Flachbart, S. Gab\'ani, S. Matas, P. Priputen, A. Efdokimova, and N. Shitsevalova, Phys. Rev. Lett. {\bf 101}, 177201 (2008).
\bibitem{TmB4_4} Z. Y. Meng and S. Wessel, Phys. Rev. B {\bf 78}, 224416 (2008).
\bibitem{KatoYasu} Y. Kato and N. Kawashima, Phys. Rev. E {\bf 79}, 021104 (2009).
\bibitem{firstorder} M. E. Fisher and A. N. Berker, Phys. Rev. B {\bf 26}, 2507 (1982).
\bibitem{Nijs} M. denNijs, Phys. Rev. B {\bf 46}, 10386 (1992).
\bibitem{AT} J. Cardy, J. Phys. A: Math. Gen. {\bf 13}, 1507 (1980). 
\bibitem{suzuki2} T. Suzuki, Y. Tomita, and N. Kawashima, (unpublished).
\bibitem{comment} $J$ is evaluated without a factor $S^2$, which is derived from the amplitude of the moment.
%
\end{references}
%%%%%%%%%%%%%%%%%%%%%%%%%%%%%%%%%%%%%%%%%%%%%%%%%%%%%%%%%%%%%%%%%%%%%%%%%%%%%

%

\end{document}